\documentclass{camera}
\usepackage{graphicx}  

\def\lesssim{\mathrel{\hbox{\rlap{\hbox{\lower4pt\hbox{$\sim$}}}\hbox{$<$}}}}
\def\gtrsim{\mathrel{\hbox{\rlap{\hbox{\lower4pt\hbox{$\sim$}}}\hbox{$>$}}}}
\newcommand{\ffrac}[2]{\left( \frac{#1}{#2} \right)}

\begin{document}

%
\title{ UPWARD AND HORIZONTAL TAU AIR-SHOWERS FROM EARTH CRUST CROWN:\\
TRACES OF HIGHEST ENERGY NEUTRINO BEAM DUMP IN WIDEST VOLUMES}

%


\author{D.Fargion}
%
\organization{Physics department, Universita' degli studi "La Sapienza", \\
              Piazzale Aldo Moro 5, -  00185 Roma, Italy\\
              INFN Roma, Istituto Nazionale di Fisica Nucleare, Italy}

\maketitle

%

\begin{abstract}
High Energy Neutrino signals from PeVs up to  GZK, $E_{\nu}\geq
10^{19}$ eV, energies  amplified as Upward and Horizontal Tau air
Showers,\textbf{UPTAUS}, and \textbf{HORTAUS}, may flash toward
mountain peaks, airplanes, balloons  and satellites in future Tau
Air-Showering experiments. Observing from  high mountains the
detectable crown masses at UHE $\nu$ energies around EeVs  are
comparable to a few $km^3$, while observing from higher balloons
or satellites at orbit altitudes, at GZK  energies  , the
Horizontal Crown effective Masses  may even exceed $150$ $km^3$.
The consequent event rate by expected GZK or Z-Showering
neutrinos models must arise at large rate by present scheduled
(EUSO-like) or future (Crown Array Detector) satellite in Space.
\end{abstract}
\maketitle
\section{INTRODUCTION: UHE $\nu$  Astronomy}
    Rarest TeVs gamma signals are at present the most extreme and
    rarest trace of  High Energy Astrophysics. The TeVs signals have shown new
power-full Jets blazing to us from Galactic or extragalactic
edges. At PeVs energies astrophysical Gamma cosmic rays should
also be present, but, excluded a very rare and elusive Cyg$X3$
event, they are  not longer being observed. While the
corresponding PeVs charged cosmic rays are abundantly hitting the
atmosphere, these missing PeVs gamma sources are very probably
mostly absorbed by their photon interactions (photo-pion
productions, electron pairs creation) at the source environment
and/or along the photon propagation into the cosmic Black Body
Radiation (BBR) or into other diffused (radio,infrared,optical)
background radiation. Unfortunately PeVs charged cosmic rays,
bend and bounded in a  random walk by Galactic magnetic fields,
loose their original directionality and their astronomical
relevance; their resident time in the galaxy is much longer ($\geq
10^{3}$-$10^{5}$) than neutral ones, as gamma rays, making the
charged cosmic rays more probable to be observed by nearly a
comparable ratio. On the contrary astrophysical UHE neutrino
signals at $10^{13}$eV-$10^{19}$eV (or higher GZK energies) are
unaffected by any radiation cosmic opacity and may open a very
new exciting window to High Energy Astrophysics. Lower energy
astrophysical UHE $\nu$ at $10^{9}$eV-$10^{12}$eV should also be
present, but their signals are (probably) drowned  by the dominant
diffused atmospheric $\nu$ secondaries noises produced by the
same charged (and smeared) UHE cosmic rays (while hitting
terrestrial atmosphere), the so called atmospheric neutrinos. In
a very far corner, at lowest (MeVs) energy  windows, the abundant
and steady solar neutrino flux  and the prompt (but rarer)
neutrino burst from a nearby Super-Novae (SN 1987A), have been in
last twenty years, already successfully explored.
\subsection{The UHE $\nu$ $Km^3$ detectors} The UHE
$10^{13}$eV-$10^{16}$eV $\nu$ 's  astronomy, being weakly
interacting and rarer, may  be captured mainly inside huge
volumes, bigger than Super-Kamiokande ones; at present  most
popular detectors consider underground ones (Cubic Kilometer Size
like AMANDA-NESTOR)\cite{Halzen2002} or (at higher energy
$10^{19}$eV-$10^{21}$eV) the widest Terrestrial atmospheric sheet
volumes (Auger-Array Telescope or EUSO atmospheric Detectors).
 Underground $km^3$ detection is based mainly on $\nu_{\mu}$ (above hundred TeVs
energies, after their interaction with matter) leading to $\mu$
kilometer size lepton tracks \cite{Gandhi et al 1998}. Rarest
atmospheric horizontal shower are also expected by $\nu$
interactions in air (and, as we shall discuss, in the Earth
Crust). While $km^3$ detectors are optimal for PeVs neutrino
muons, the Atmospheric Detectors (AUGER-EUSO like) exhibit a
minimal threshold at highest ($\geq 10^{18} eV$) energies.
 \subsection{ UHECR EeV $n$ and PeV$\nu$  Astronomy}
 Incidentally just around such EeV ($10^{18} eV$) energies
an associated Ultra High Energy Neutron Astronomy  might be
possible because of the relativistic neutrons boosted lifetime,
comparable to our distance from the Galactic center. Therefore
UHE  neutrons at EeV may be source of part of UHECR data. Indeed a
$4\%$ galactic anisotropy and clustering in EeV cosmic rays is
recently emerging \cite{AGASA 1999} along our nearby galactic
plane. Therefore AGASA might have already experienced a first
UHECR-Neutron astronomy \textbf{UHENA}. This EeV-UHENA signals
must also be source of a parasite ($10^{17}-10^{16}$ eV)
secondary tails of UHE neutrino $\bar{\nu_{e}}$  from the same
neutron beta decay in flight. Their oscillations in galactic or
extragalactic flights (analogous to atmospheric and solar ones)
must guarantee the presence of all lepton flavours nearly at
equal foot: $\bar{\nu_{e}}$,$\bar{\nu_{\mu}}$ $\bar{\nu_{\tau}}$.
 The latter UHE $\bar{\nu_{\tau}}$ imprint (added to other local astrophysical UHE $\nu$
production) could be already recorded \cite{Fargion2000-2002} as
Upward Tau Air-Showers  Terrestrial Gamma Flash (assumed
secondaries $\gamma$ of Upward Tau Air-Showers and Horizontal Tau
Air-Showers). These estimates are part of present article below.
 \subsection{ GZK $\nu$ Astronomy and Z-Showering}
 Finally,  let us also remind that at highest energy edges ($\geq
10^{19}-10^{20} eV$), a somehow correlated New UHE Astronomy is
also expected for charged Cosmic Rays; indeed these UHECR have
such a large rigidity to avoid any bending by random galactic or
extragalactic magnetic fields;  being nearly undeflected UHECR
should point toward the original sources showing in sky a new
astronomical map. (There have been very reiterated attempts to
consider extreme galactic or extragalactic coherent magnetic
fields able to bend the well known nearby, like Virgo Clusters, M
$87$ AGN, sources in a diffused map, however their needed magnetic
energy density and coherent lenghts are in our opinion totally
unrealistic). Moreover such UHECR astronomy is bounded by the
ubiquitous cosmic $2.75 K^{o}$ BBR screening (the well known
Greisen,Zat'sepin,Kuzmin GZK cut-off) limiting its origination
inside a very local ($\leq 20 Mpc$) cosmic volume. Surprisingly,
these UHECR above GZK (already up to day above $60$ events) are
not pointing toward any known nearby candidate source. Moreover
their nearly isotropic arrival distributions underlines and
testify a very possible cosmic origination, in disagreement with
any local (Galactic plane or Halo) expected footprint by GZK
cut-off. A very weak Super-Galactic imprint seem to be present
but already above GZK volume.  This opened a very hot debate in
modern astrophysics known as the GZK paradox. Possible solutions
has been found recently beyond Standard Model assuming a
non-vanishing neutrino mass. Indeed at such Ultra-High energies,
neutrino at ZeV energies ($\geq 10^{21}eVs$) hitting onto  relic
cosmological light ($0.1-4 eV$ masses) neutrinos
\cite{Dolgov2002} nearly at rest in Dark Hot Halos (galactic or
in Local Group) has the unique possibility to produce UHE
resonant Z bosons (the so called Z-burst or better Z-Showering
scenario): \cite{Fargion Salis 1997}, \cite{Fargion Mele  Salis
1999}, \cite{Yoshida  et all 1998},\cite{Weiler 1999}; for a more
updated scenario see \cite{Fargion et all. 2001b}. Indeed UHE
neutrinos are un-effected by magnetic fields and by BBR
screening; they may reach us from far cosmic edges with
negligible absorption. The UHE Z-shower in its ultra-high energy
nucleonic secondary component may be just the observed final
UHECR event on Earth. This possibility has been reinforced by
very recent correlations (doublets and triplets events) between
UHECR directions with brightest Blazars sources at cosmic
distances (redshift $\geq 0.1$) quite beyond ($\geq 300 Mpc$) any
allowed GZK cut-off \cite{Tinyakov-Tkachev2001} \cite{AGASA
1999},\cite{Takeda et all}. Therefore there might be a role for
GZK neutrino fluxes, either as primary in the Z-Showering
scenario or, at least, as (but at lower intensities) necessary
secondaries of all those UHECR primary absorbed in cosmic BBR
radiation fields. Naturally other solutions as topological
defects or primordial relics decay may play a role as a source of
UHECR, but the observed clustering \cite{AGASA 1999},
\cite{Tinyakov-Tkachev2001},\cite{Takeda et all}  seems to favor
compact sources. The most recent evidence for self-correlations
clustering at $10^{19}$, $2\cdot10^{19}$, $4\cdot10^{19}$ eVs
energies observed by AGASA (Teshima, ICRR26 Hamburg presentation
2001) maybe a first reflection of UHECR Z-Showering secondaries:
$p, \bar{p}, n, \bar{n}$ \cite{Fargion et all. 2001b}. A very
recent solution beyond the Standard Model (but within
Super-Symmetry) consider Ultra High Energy Gluinos as the neutral
particle bearing UHE signals interacting nearly as an hadron in
Terrestrial Atmosphere \cite{Berezinsky 2002}; this solution has
a narrow window  for gluino masses allowable (and serious
problems in production bounds), but it is an alternative that
deserves attention. To conclude the puzzle one finally needs to
scrutiny the UHE $\nu$ astronomy and to test the GZK solution
within Z-Showering Models by any independent search on Earth for
such UHE neutrinos traces above PeVs reaching either EeVs-ZeVs
extreme energies.
 \subsection{UHE $\nu$ Astronomy by $\tau$ Air-Shower}
 Recently \cite{Fargion et all 1999},\cite{Fargion2000-2002}
 it has been proposed a new competitive (and in our opinion,
more convenient) UHE $\nu$ detection based on ultra high energy
$\nu_{\tau}$ interaction in matter and its consequent secondary
$\tau$ decay in flight while escaping from the rock (Mountain
Chains) or water (Sea)  in air leading to Upward or Horizontal
$\tau$ Air-Showers (UPTAUs and
HORTAUs),\cite{Fargion2001a},\cite{Fargion2001b}.  In a pictorial
way one may compare the UPTAUs and HORTAUs as the double bang
processes expected in $km^3$ ice-water volumes \cite{Learned
Pakvasa 1995} : the double bang is due first to the UHE
$\nu_{\tau}$ interaction in matter and secondly by its consequent
$\tau$ decay in flight. Here we consider  a (hidden) UHE $\nu$-N
Bang $in$ (the rock-water within a mountain or the Earth Crust)
and a $\tau$ bang $out$ in air, whose shower is better observable
at high altitudes. The main power of the UPTAUs and HORTAUs
detection is the huge amplification of the UHE neutrino signal,
which may deliver almost all its energy in numerous secondaries
traces (Cherenkov lights, gamma, X photons, electron pairs,
collimated muon bundles). Indeed the multiplicity in $\tau$
Air-showers secondary particles, $N_{opt} \simeq 10^{12}
(E_{\tau} / PeV)$, $ N_{\gamma} (< E_{\gamma} > \sim  10 \, MeV )
\simeq 10^8 (E_{\tau} / PeV) $ , $N_{e^- e^+} \simeq 2 \cdot 10^7
( E_{\tau}/PeV) $ , $N_{\mu} \simeq 3 \cdot 10^5
(E_{\tau}/PeV)^{0.85}$ makes easy the UPTAUs-HORTAUs discover.
 These HORTAUs, also named Skimming neutrinos \cite{Feng et al 2002},
 studied also in peculiar approximation in the frame of AUGER experiment,
 \cite{Bertou et all 2002}, maybe also
originated on front of mountain chains \cite{Fargion et all
1999}, \cite{Hou Huang 2002}  either by $\nu_{\tau}N$, $
\bar\nu_{\tau}N$ interactions as well as by $ \bar\nu_{e} e
\rightarrow W^{-} \rightarrow \bar\nu_{\tau} \tau$. This new UHE
$\nu_{\tau}$ detection is mainly based on the oscillated UHE
neutrino $\nu_{\tau}$ originated by more common astrophysical
$\nu_{\mu}$, secondaries of pion-muon decay at PeVs-EeVs-GZK
energies. These oscillations are guaranteed  by Super Kamiokande
evidences for flavour mixing within GeVs atmospheric neutrino
data as well as by most solid and recent evidences of complete
solar neutrino mixing observed by SNO detector. HORTAUs from
mountain chains must nevertheless occur, even for no flavour
mixing, as being inevitable $\bar\nu_{e}$ secondaries of common
pion-muon decay chains ($\pi^{-}\rightarrow
\mu^{-}+\bar\nu_{\mu}\rightarrow e^{-}+\bar\nu_{e}$) near the
astrophysical sources at Pevs energies. They are mostly absorbed
by the Earth and are only rarely arising as UPTAUS. Their Glashow
resonant interaction allow them to be observed as HORTAUs only
within a very narrow and nearby crown edges at horizons (not to be
discussed here). At wider energies windows ($10^{14}eV-
10^{20}eV$) only neutrino $\nu_{\tau}$, $\bar{\nu}_{\tau}$ play a
key role in UPTAUS and HORTAUS. These Showers might be easily
detectable looking downward the Earth's surface from mountains,
planes, balloons or satellites observer. Here the Earth itself
acts as a "big mountain" or a wide beam dump target. The present
upward $\tau$ at horizons should not be confused with an
independent and well known, complementary (but rarer) Horizontal
Tau Air-shower originated inside the same terrestrial atmosphere:
we shall referee to it as the Atmospheric Horizontal Tau
Air-Shower. The same UPTAUS have a less competitive upward
showering due to $\nu_{e}$ $\bar\nu_{e}$ interactions with
atmosphere, showering in thin upward air layers \cite{Berezinsky
1990}: let us label this atmospheric Upward Tau as A-UPTAUs  and
consider its presence  as a very small additional contribute,
because rock is more than $3000$ times denser than air. Therefore
at different heights we need to estimate the UPTAUS and HORTAUs
event rate occurring along the thin terrestrial crust below the
observer, keeping care of their correlated  variables: from a
very complex sequence of functions we shall be able to define and
evaluate the effective HORTAUs volumes keeping care of the thin
shower beaming angle, atmosphere opacity and detector thresholds.
At the end of the study, assuming any given neutrino flux, one
might be easily able to estimate at each height $h_{1}$ the
expected event rate and the ideal detector size and sensibility
for most detection techniques (Cherenkov, photo-luminescent,
gamma rays, X-ray, muon bundles).
\section{THE  UPTAUs-HORTAUs  detection}
The $\tau$ airshowers are observable at different height $h_{1}$
 leading to different underneath observable terrestrial
areas and crust volumes. HORTAUs in deep valley are also relate
to the peculiar geographical morphology and composition
\cite{Fargion2000-2002} and more in detail as discussed below. We
remind in this case the very important role of UHE  $ \bar\nu_{e}e
\rightarrow W^{-}\rightarrow \bar\nu_{\tau}\tau^{-} $ channels
which may be well observable even in absence of any $\nu_{\tau}$,
$ \bar\nu_{\tau}$ UHE sources or any neutrino flavour mixing: its
Glashow peak resonance make these neutrinos unable to cross all
the Earth across but it may be observable beyond mountain chain
\cite{Fargion2000-2002}; while testing $\tau$ air-showers beyond
a mountain chain one must keep in  mind the possible amplification
of the signal because of a possible New TeV Physics (see Fig 5)
\cite{Fargion2000-2002}. In the following we shall consider in
general the main $\nu_{\tau}-N$,$ \bar\nu_{\tau}-N$ nuclear
interaction on Earth crust. It should be kept in mind also that
UPTAUs and in particular HORTAUS are showering at very low
densities and their geometrical opening angle (here assumed at
$\theta\sim 1^o$) is not in general conical (like down-ward
showers) but they are more in a thin fan-like shape (like the
observed  $8$ shaped horizontal Air-Showers). The fan shape is
opened by the Terrestrial magnetic field bending. These
UPTAUs-HORTAUs duration time is also much longer than common
down-ward showers because their showering occurs at much lower air
density: from micro (UPTAUS from mountains) to millisecond (UPTAUs
and HORTAUs from satellites) long flashes. Indeed the GRO
observed Terrestrial Gamma Flashes, possibly correlated with the
UPTAUs \cite{Fargion2000-2002} show the millisecond duration
times. In order to estimate the rate and the fluence for of
UPTAUs and HORTAUs one has to  estimate the observable mass,
facing a complex chain of questions, leading for each height
$h_{1}$, to an effective observable surface and volume from where
UPTAUs and HORTAUs might be originated. From this effective
volume it is easy to estimate the observable rates, assuming a
given incoming UHE $\nu$ flux model for galactic or extragalactic
sources. Here we shall only refer to the Masses
estimate,unrelated to any UHE $\nu$ flux models. These steps are
linking simple terrestrial spherical geometry and its different
geological composition, high energy neutrino physics and UHE
$\tau$ interactions, the same UHE $\tau$ decay in flight and its
air-showering physics at different quota within terrestrial air
density. Detector physics threshold and background noises, signal
rates have been kept in mind \cite{Fargion2000-2002}, but they
will be discussed and explained in  forthcoming papers.
\section{The Skin  Crown Earth Volumes }
Let us therefore define, list and estimate below the sequence of
the key variables whose dependence (shown below or derived in
Appendices) leads to the desired HORTAUs volumes (useful to
estimate the UHE $\nu$ prediction rates) summirized in a last
Table and in Conclusions. These Masses estimate are somehow an
lower bound that ignore additional contribute by more penetrating
or regenerated $\tau$. \cite{Halzen1998}.  Let us now show the
main functions whose interdependence with the observer altitude
lead to estimate the UPTAUs and HORTAUs equivalent detection
Surfaces, Volumes and Masses.
\begin{enumerate}
  \item The horizontal distance $d_{h}$ at given height $h_{1}$
  toward the horizons:

 \begin{equation}
 d_{h}= \sqrt{( R_{\oplus} + h_1)^2 - (R_{\oplus})^2}= 113\sqrt{ \frac{h_1}{km} }\sqrt{1+\frac{h_1}{2R_{\oplus}} }\cdot {km}
 \end{equation}

The corresponding horizontal edge angle $\theta_{h}$ below the
horizons ($\pi{/2}$) is:

\begin{equation}
 {\theta_{h} }={\arccos {\frac {R_{\oplus}}{( R_{\oplus} + h_1)}}}\simeq 1^o \sqrt{\frac {h_{1}}{km}}
\end{equation}

(All the approximations here and below hold for height
$h_{1}\ll{R_{\oplus}}$ )

\begin{figure}\centering\includegraphics[width=11cm]{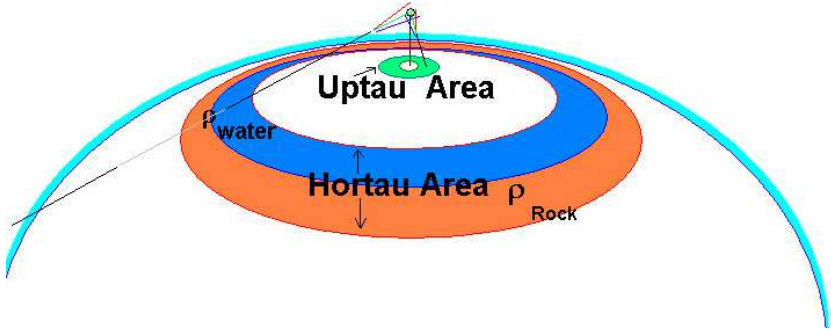}
\caption {The Upward Tau Air-Shower, UPTAUs,and the Horizontal
Tau Air-Shower, HORTAU, flashing toward an observer at height
$h_1$. The HORTAU areas are described for water and rock matter
density. } \label{fig:fig1}
\end{figure}
 \begin{figure}\centering\includegraphics[width=11cm]{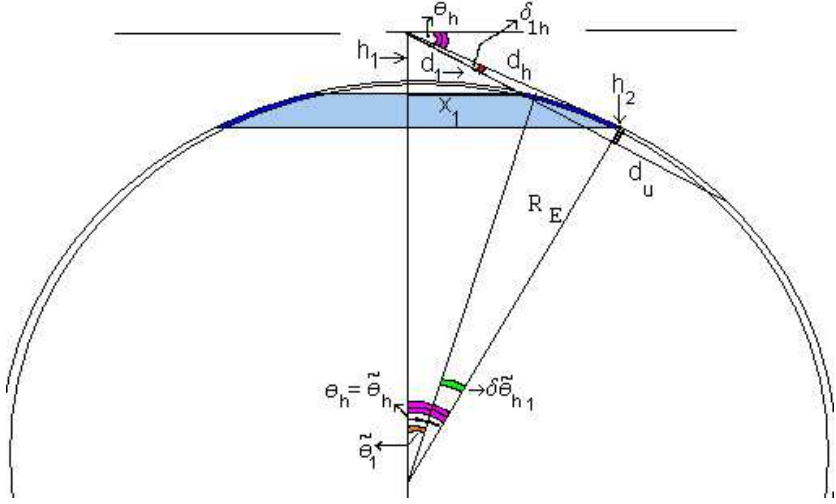}
\caption {The geometrical disposal of the main parameters in the
text defining the UPTAUs and HORTAUs Areas; the distances are
exaggerated for simplicity.} \label{fig:fig2}
\end{figure}
  \item
  The consequent characteristic lepton $\tau$ energy
  $E_{\tau_{h}}$ making decay  $\tau$ in flight from  $d_{h}$ distance just nearby
  the source:

\begin{displaymath}
 {E_{\tau_{h}}}=  {\left(\frac {d_{h}}{c\tau_{0}}\right)}m_{\tau} c^2=
\end{displaymath}

\begin{equation}
\simeq{2.2\cdot10^{18}eV}\sqrt{\frac {h_{1}}{km}}\sqrt{1 + \frac
{h_{1}}{2R}}
\end{equation}
At low quota ($h_1 \leq$ a few kms) the  air depth before the Tau
decay necessary to develop a shower correspond to a Shower
distance $d_{Sh}$ $\sim 6 kms \ll d_h$.  More precisely at low
quota ($h_1\ll h_o$, where $h_o$ is the air density decay height$
= 8.55$ km.) one finds:
\begin{equation}
d_{Sh}\simeq 5.96km[ 1+ \ln {\frac{E_{\tau}}{10^{18}eV}})] \cdot
e^{\frac{h_1}{h_o}}
\end{equation}
  So we may neglect the distance of the final shower respect to the
longest horizons ones. However at high  altitude ($h_1\geq h_o$)
this is no longer the case (see Appendix A). Therefore  we shall
introduce from here and in next steps a small, but important
modification , whose physical motivation is just to include  the
air dilution role at highest quota: $ {h}_1 \rightarrow \frac
{h_{1}}{1 + h_1/H_o} $, where , as in Appendix A, $H_o= 23$ km.
Therefore previous definition becomes:
\begin{equation}
 {E_{\tau_{h}}}\simeq{2.2\cdot10^{18}eV}\sqrt{\frac {h_{1}}{1 + h_1/H_o}}\sqrt{1 + \frac
{h_{1}}{2R}}
\end{equation}
This procedure, applied tacitly everywhere, guarantees that there
we may extend our results to those HORTAUs at altitudes where the
residual air density  must exhibit a sufficient slant depth. For
instance, highest $\gg 10^{19} eV$ HORTAUs will be not easily
observable because their ${\tau}$ life distance exceed (usually)
the horizons air depth lenghts.
\item
The parental UHE $\nu_{\tau}$,$\bar\nu_{\tau}$ or $\bar\nu_{e}$
energies $E_{\nu_{\tau}}$ able to produce such UHE $E_{\tau}$ in
matter:
\begin{equation}
 E_{\nu_{\tau}}\simeq 1.2 {E_{\tau_{h}}}\simeq {2.64
 \cdot 10^{18}eV \cdot \sqrt{\frac {h_{1}}{km}}}
\end{equation}
\item
The neutrino (underground) interaction lenghts  at the
corresponding energies is $L_{\nu_{\tau}}$:
\begin{displaymath}
L_{\nu_{\tau}}= \frac{1}{\sigma_{E\nu_{\tau}}\cdot N_A\cdot\rho_r}
\end{displaymath}
\begin{displaymath}
= 2.6\cdot10^{3}km\cdot \rho_r^{-1}{\left(\frac{E_{\nu_h}}{10^8
\cdot GeV}\right)^{-0.363}}
\end{displaymath}
\begin{equation}
{\simeq 304 km \cdot
\left(\frac{\rho_{rock}}{\rho_r}\right)}\cdot{\left(\frac{h_1}{km}\right)^{-0.1815}}
\end{equation}
 For more details see \cite{Gandhi et al 1998}, \cite{Fargion2000-2002}.
 It should be remind that here we ignore the $\tau$ multi-bangs \cite{Halzen1998}
 that reduce the primary $\nu_{\tau}$ energy and pile up the lower
 energies HORTAUs (EeV-PeVs).

\item
The maximal neutrino depth $h_{2}(h_{1})$ under the chord along
the UHE neutrino-tau trajectory of lenght $L_{\nu}(h_{1})$:
\begin{displaymath}
h_{2}(h_{1}) = {\frac{L_{\nu_{h}}^2}{2^2\cdot{2}(R-h_{2})}
\simeq\frac{L_{\nu_{h}}^2}{8R_{\oplus}}}\simeq
\end{displaymath}
\begin{equation}
{\simeq 1.81\cdot km
\cdot{\left(\frac{h_1}{km}\right)^{-0.363}\cdot
\left(\frac{\rho_{rock}}{\rho_r}\right)^2}}
\end{equation}
See figure 2, for more details. Because the above $h_2$ depths
are in general not  too deep respect to the Ocean depths, we shall
consider either sea (water) or rock (ground) materials as Crown
matter density.
\item
The corresponding opening angle observed from height $h_{1}$,
$\delta_{1h}$ encompassing the underground height  $h_{2}$ at
horizons edge (see Fig.2) and the nearest UHE $\nu$ arrival
directions $\delta_{1}$:
\begin{displaymath}
{{\delta_{1h}}(h_{2})}={2\arctan{\frac{h_{2}}{2 d_{h}}}}=
\end{displaymath}
\begin{displaymath}
={2\arctan\left[\frac{{8\cdot
10^{-3}}\cdot{(\frac{h_{1}}{km}})^{-0.863}\left(\frac{\rho_{rock}}{\rho_r}\right)^2}{{\sqrt{1+{\frac{h_{1}}{2R}}}}}\right]}
\end{displaymath}
\begin{equation}
{\simeq 0.91^{o}
\left(\frac{\rho_{rock}}{\rho_r}\right)^2}\cdot{(\frac{h_{1}}{km}})^{-0.863}
\end{equation}
\item
The underground chord $d_{u_{1}}$ (see Fig.$2-4$) where UHE
$\nu_{\tau}$ propagate and the nearest distance $d_{1}$ for
$\tau$ flight (from the observer toward Earth) along the same
$d_{u_{1}}$ direction, within the angle $\delta_{1h}$ defined
above, angle below the horizons (within the upward UHE neutrino
and HORTAUs propagation line) is:
\begin{equation}
d_{u_{1}}=2\cdot{\sqrt{{\sin}^{2}(\theta_{h}+\delta_{1h})(R_{\oplus}+{h_{1}})^{2}-{d_{h}}^2}}
\end{equation}
 Note that by definition  and by construction:
 \begin{equation}
 L_{\nu} \equiv d_{u_{1}}
\end{equation}
The nearest HORTAUs distance corresponding to this horizontal
edges still transparent to UHE $\tau$ is:
\begin{equation}
{d_{1}(h_{1})}=(R_{\oplus}+h_{1})\sin(\theta_{h}+\delta_{1h})-{\frac{1}{2}}d_{u_{1}}
\end{equation}

 Note also that for height $h_{1}\geq km$ :
\begin{displaymath}
\frac{d_{u_{1}}}{2}\simeq{(R_{\oplus}+{h_{1}})\sqrt{\delta_{1h}\sin{2\theta_{h}}}}\simeq
\end{displaymath}

\begin{equation}
\simeq{158\sqrt{\frac{\delta_{1h}}{1^o}}\sqrt{\frac{h_{1}}{km}}}km
\end{equation}
\begin{figure}\centering\includegraphics[width=11cm]{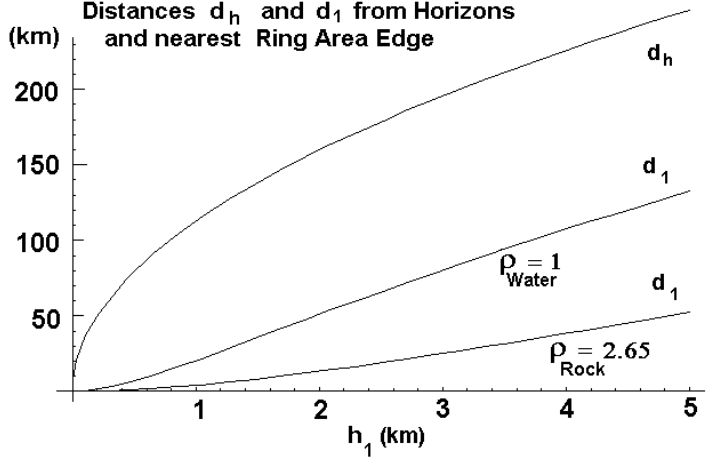}
\caption {Distances from the observer to the Earth ($d_1$) for
different matter densities or to the Horizons ($d_h$) for low
altitudes} \label{fig:fig3}
\end{figure}

\item
The same distance projected cord $x_{1}(h_{1})$ along the
horizontal line (see Fig.2):
\begin{equation}
x_{1}(h_{1})=d_{1}(h_{1})\cos({\theta_{h}+\delta_{1h}})
\end{equation}
\item
The total terrestrial underneath any observer at height $h_{1}$
is $A_{T}$:
\begin{displaymath}
=2\pi{R_{\oplus}}^{2}(1-\cos{\tilde{\theta}_{h}})
=2\pi{R_{\oplus}}h_{1}\left({\frac{1}{1+\frac{h_{1}}{R_{\oplus}}}}\right)
\end{displaymath}
\begin{equation}
A_{T}=4\cdot{10}^{4}{km}^{2}{\left({\frac{h_{1}}{km}}\right)}{\left({\frac{1}{1+{\frac{h_{1}}{R}}}}\right)}
\end{equation}
Where $\tilde{\theta}_{h}$ is the opening angle from the Earth
along the observer and the horizontal point whose value is the
maximal observable one. At first sight one may be tempted to
consider all the Area  $A_{T}$ for UPTAUs and HORTAUs but because
of the air opacity (HORTAUs) or for its paucity (UPTAUs) this is
incorrect.  While for HORTAUs there is a more complex Area
estimated above and in the following, for UPTAUs the Area Ring (or
Disk) is quite simpler to derive following very similar
geometrical variables summirized in Appendix B.
\item
The Earth Ring Crown crust area ${A_{R}}(h_{1})$ delimited by the
horizons distance $d_{h}$ and the nearest distance $d_{1}$ still
transparent to UHE $\nu_{\tau}$. The ring area ${A_{R}}(h_{1})$
is computed from the internal angles $\delta{\tilde{\theta}_{h}}$
and $\delta{\tilde{\theta}_{1}}$ defined at the Earth center
(note that $\delta{\tilde{\theta}_{h}}={\delta{\theta_{h}}}$ but
in general $\delta{\tilde{\theta}_{1}}\neq{\delta{\theta_{1}}}$).

\begin{equation}
{A_{R}}(h_{1})=2\pi{R_{\oplus}}^2(\cos{\tilde{\theta}_{1}}-\cos{\tilde\theta_{h}})
\end{equation}

\begin{equation}
=2\pi{R_{\oplus}^{2}}{\left({{\sqrt{1-{\left({\frac{x_{1}({h_1})}{R_{\oplus}}}\right)^{2}}}}-{\frac{R_{\oplus}}{R_{\oplus}+{h_{1}}}}}\right)}
\end{equation}
 Here $x_{1}({h_1})$ is the cord defined above.


\begin{figure}\centering\includegraphics[width=11cm]{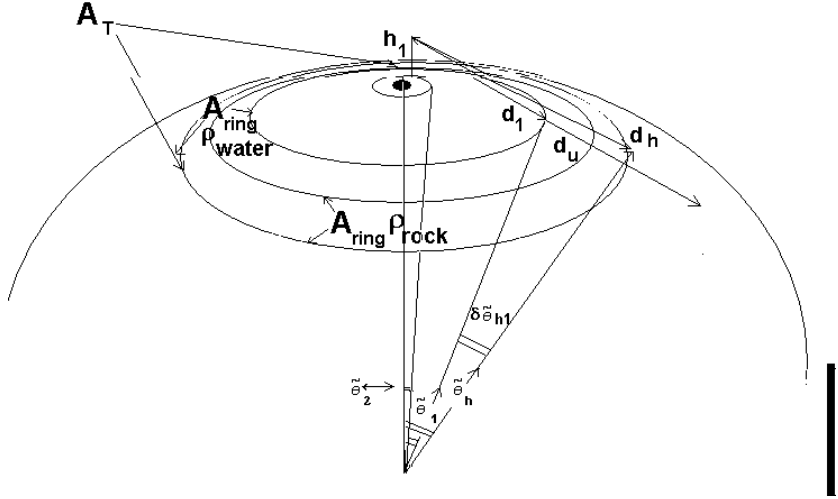}
\caption {Areas and Angles for UPTAUS-HORTAUS } \label{fig:fig4}
\end{figure}
\begin{figure}\centering\includegraphics[width=11cm]{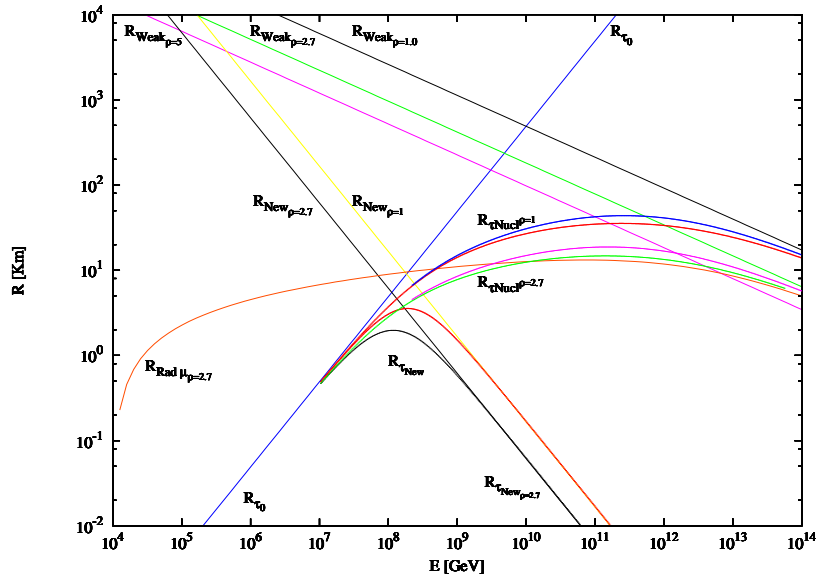}
\caption {Lepton $\tau$ (and $\mu$) Interaction Lenghts for
different matter density: $R_{\tau_{o}}$ is the free $\tau$
lenght,$R_{\tau_{New}}$ is the New Physics TeV Gravity
interaction  range at corresponding
densities,$R_{\tau_{Nucl}\cdot{\rho}}$ ,\cite{Fargion2000-2002},
see also \cite{Becattini Bottai 2001}, \cite{Dutta et al.2001}, is
the combined $\tau$ Ranges keeping care of all known interactions
and lifetime and mainly the photo-nuclear one. There are two
slightly different split curves (for each density) by two
comparable approximations in the interaction laws.
$R_{Weak{\rho}}$ is the electro-weak Range at corresponding
densities (see also \cite{Gandhi et al
1998});\cite{Fargion2000-2002}.} \label{fig:fig5}
\end{figure}
\item
The characteristic interaction lepton tau lenght $l_{\tau}$
defined at the average $E_{\tau_{1}}$, from interaction in matter
(rock or water). These lenghts have been derived by a analytical
equations keeping care of the Tau lifetime, the photo-nuclear
losses, the electro-weak losses \cite{Fargion2000-2002}. See
figure 5 below.

\item
The $l_{\tau_{2}}$ projected along the
$\sin(\delta\tilde{\theta}_{h_{1}})$:
\begin{equation}
\delta\tilde{\theta}_{h_{1}}\equiv \tilde\theta_{h}
-\arcsin{\left({\frac{x_{1}}{R_{\oplus}}}\right)}
\end{equation}

\begin{figure}\centering\includegraphics[width=8cm]{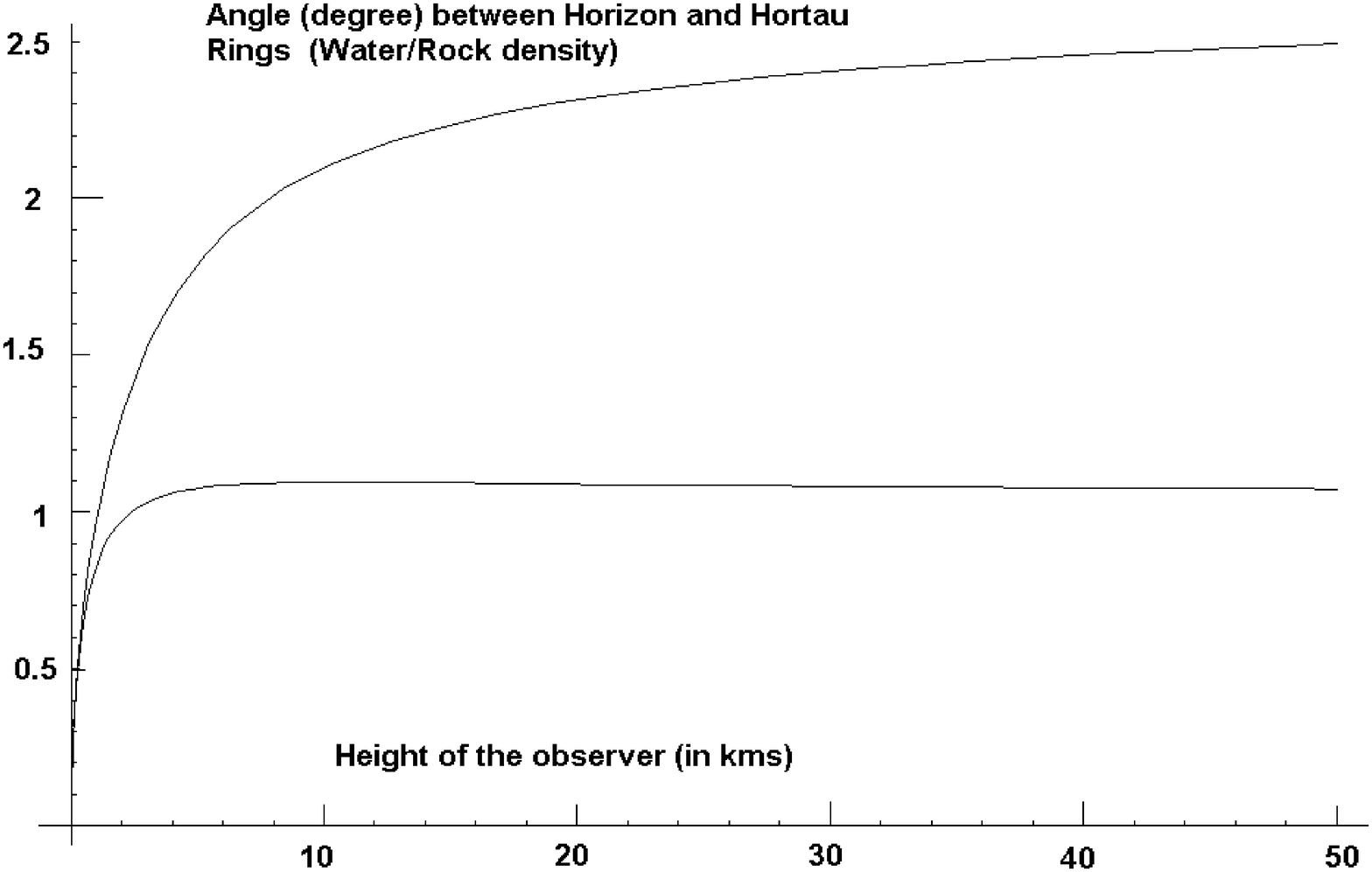}
\caption {The $\delta\tilde{\theta}_{h_{1}}$ opening angle toward
Ring Earth Skin for density $\rho_{water}$  and $\rho_{rock}$ }
\label{fig:fig6}
\end{figure}
The same quantity in a more direct approximation:
\begin{displaymath}
\sin\delta{\tilde{\theta}_{h_{1}}}\simeq\frac{L_{\nu}}{2R_{\oplus}}=\frac{{304}km}{2R_{\oplus}}{\left({\frac{\rho_{rock}}{\rho}}\right)}{\frac{h_{1}}{km}}^{-0.1815}.
\end{displaymath}
From highest ($h\gg H_o$=23km) altitude the exact approximation
reduces to:
$$\delta{\tilde{\theta}_{h_{1}}}\simeq{1}^o{\left({\frac{\rho_{rock}}{\rho}}\right)}\left({\frac{h_{1}}{500\cdot
km}}\right)^{-0.1815}$$ Therefore the penetrating $\tau$ skin
depth $l_{\tau_{\downarrow}}$ is
\begin{equation}
l_{\tau_{\downarrow}}=l_{\tau}\cdot\sin\delta{\tilde{\theta}_{h_{1}}}
\simeq{{0.0462\cdot
l_{\tau}{\left({\frac{\rho_{water}}{\rho}}\right)}}}{\frac{h_{1}}{km}}^{-0.1815}
\end{equation}

Where the $\tau$ ranges in matter, $l_{\tau}$ has been calculated
and shown in Fig.5.


\begin{figure}\centering\includegraphics[width=13cm]{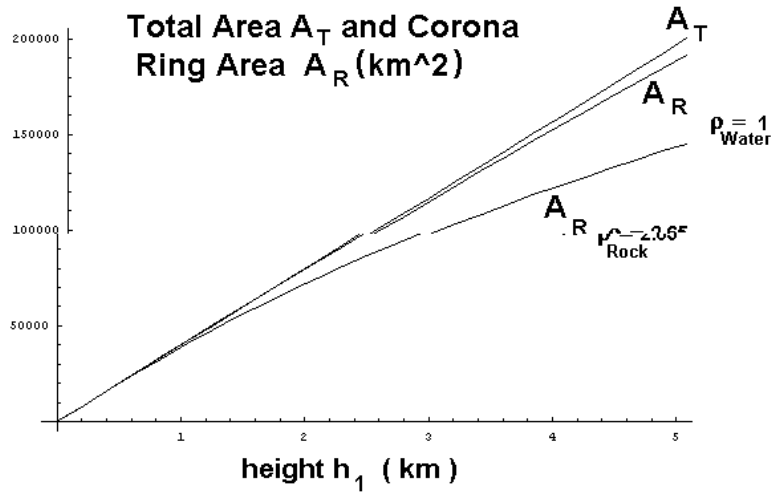}
\caption {Total Area $A_T$ and Ring Areas for two densities $A_R$
at low altitudes }\label{fig:fig7}
\end{figure}
\begin{figure}\centering\includegraphics[width=13cm]{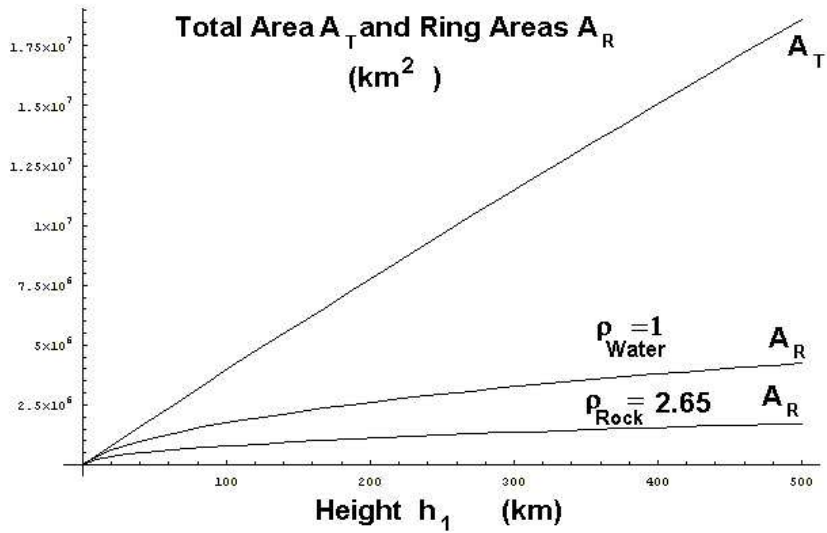}
\caption {Total Area $A_T$ and Ring Areas for two densities $A_R$
at high altitudes } \label{fig:fi8}
\end{figure}

\item
The final  analytical expression for the Earth Crust Skin Volumes
and Masses under the Earth Skin inspected by HORTAUs are derived
combining the above functions on HORTAUs Areas  with the previous
lepton Tau $l_{\tau_{\downarrow}}$ vertical depth depths:
\begin{equation}
{V_{h_{1}}}={A_{R}}(h_{1})\cdot l_{\tau_{\downarrow}};
\end{equation}
\begin{equation}
{M_{h_{1}}}={V_{h_{1}}}\cdot{\left({\frac{\rho}{\rho_{water}}}\right)}
\end{equation}

\item A More approximated but easy to handle
 expression for Ring area for high altitudes ($h_1\gg 2km$ $h_1\ll R_{\oplus}$) may be
summirized as:
\begin{displaymath}
{A_R (h_1)}\simeq
2\pi{R_{\oplus}^2}\sin{\theta_{h}}{\delta{\tilde{\theta}_{{h}_{1}}}}\propto{\rho^{-1}}
\end{displaymath}
\begin{equation}
\simeq{2\pi{R_{\oplus}^{2}}{\sqrt{\frac{2h_{1}}{R_{\oplus}}}}\left({\frac{\sqrt{1+{\frac{h_{1}}{2R_{\oplus}}}}}{1+{\frac{h_{1}}{R}}}}\right)}{\left({\frac{L_{\nu}}{2R_{\oplus}}}\right)}
\end{equation}
 At high altitudes the above approximation corrected
accordingly to the exact one shown in Figure, becomes:
\begin{displaymath}
{A_R (h_1)}\simeq
2\pi{R_{\oplus}}{d_{h1}}{\delta{\tilde{\theta}_{{h}_{1}}}}\simeq
\end{displaymath}
\begin{equation}
\simeq{4.65 \cdot 10^6{\sqrt {\frac{h_{1}}{500
km}}}}{\left({\frac{\rho_{water}}{\rho}}\right)}{km}^2
\end{equation}
 Within the above
approximation the final searched Volume ${V_{h_{1}}}$ and Mass
${M_{h_{1}}}$ from where HORTAUs may be generated  is:
\begin{equation}
{V_{h_{1}}}={\frac{\pi}{2}{\sqrt{\frac{2h_{1}}{R_{\oplus}}}}
\left({\frac{\sqrt{1+{\frac{h_{1}}{2R}}}}{1+{\frac{h_{1}}{R}}}}\right)}
{L_{\nu}^{2}}{l_{\tau}}\propto{\rho^{-3}}
\end{equation}
\begin{equation}
{M_{h_{1}}}={\frac{\pi}{2}{\sqrt{\frac{2h_{1}}{R_{\oplus}}}}\left({\frac{\sqrt{1+{\frac{h_{1}}{2R}}}}{1+{\frac{h_{1}}{R}}}}\right)}{L_{\nu}^{2}}{l_{\tau}}{\rho}\propto{\rho^{-2}}
\end{equation}
\item
The effective observable Skin Tau Mass $M_{eff.}(h_{1})$ within
the thin HORTAU or UPTAUs Shower angle beam $\simeq$ $1^o$ is
suppressed by the solid angle of view
${\frac{\delta\Omega}{\Omega}} \simeq 2.5\cdot 10^{-5}$.
\begin{equation}
{\Delta
M_{eff.}(h_{1})={V_{h_{1}}}\cdot{\left({\frac{\rho}{\rho_{water}}}\right){\frac{\delta\Omega}{\Omega}}}}
\end{equation}
The Masses $M_{eff.}(h_{1})$ for realistic high quota experiment
are discussed in the Table below.

\onecolumn
\begin{figure}[H]
\includegraphics[height=1.2\textwidth,angle=270]{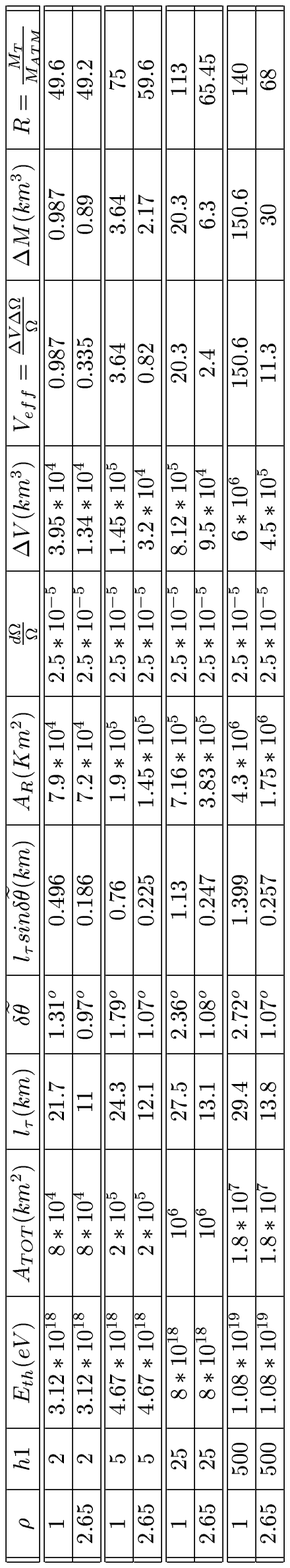}
\caption{The Table of the main parameters leading to the
effective HORTAUs Mass  from the observer height $h_1$, the
corresponding $\tau$ energy $E_{\tau}$ able to let the $\tau$
reach him from the horizons, the Total Area $A_{TOT}$ underneath
the observer, the corresponding $\tau$ propagation lenght in
matter $l_{\tau}$, the opening angle toward the Crown from the
Earth $\delta\tilde{\theta}_{h_1}$
 and  $l_{\tau}$ just orthogonal in the matter $l_{\tau_{\downarrow}}=l_{\tau}\cdot\sin\delta{\tilde{\theta}_{h_{1}}}$, the Ring Areas for
two densities $A_R$ at characteristic high altitudes $h_1$, the
corresponding effective Volume $V_{eff.}$ and the consequent Mass
$\Delta M_{eff.}$ (within the narrow $\tau$ Air-Shower solid
angle) as a function of density $\rho$ and height$h_1$. In the
last Column the Ratio R $= M_T/M_{ATM}$ define the ratio of
HORTAUs produced within the Earth Crown Skin over the atmospheric
ones: this ratio nearly reflects the matter over air density and
it reaches nearly two order of magnitude.}
\end{figure}
\vspace{-1 cm}

\begin{figure}\centering\includegraphics[width=16cm]{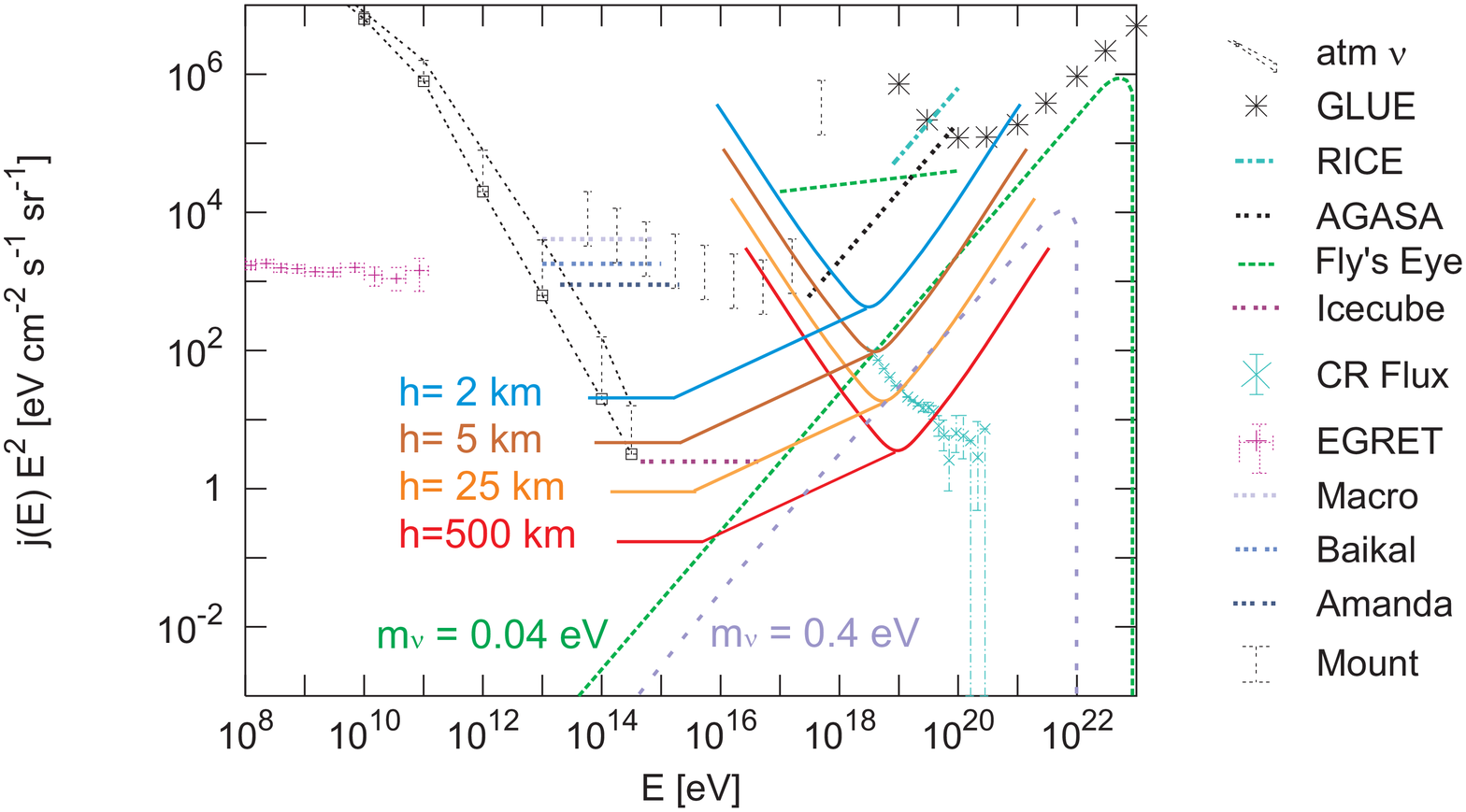}
\caption {UPTAUS (lower bound on the center) and HORTAUS (right
parabolic  curves)  sensibility at different observer heights h
($2,5,25,500 km $) assuming a $km^3$ scale volume (see Table
above)  adapted over a present neutrino flux estimate in Z-Shower
model scenario \cite{Kalashev:2002kx}, \cite{Fargion Mele Salis
1999} for light ($0.4-0.04$ eV) neutrino masses $m_{\nu}$; two
corresponding density contrast has been assumed \cite{Fargion et
all. 2001b}; the lower parabolic bound thresholds are at different
operation height, in Horizontal (Crown) Detector facing toward
most distant horizons edge; these limits are fine tuned (as
discussed in the text); we are assuming a duration of data
records of a decade comparable to the BATSE record data (a
decade). The paraboloid bounds on the EeV energy range in the
right sides  are nearly un-screened by the Earth opacity while
the corresponding UPTAUS bounds  in the  center below suffer both
of Earth opacity as well as of a consequent shorter Tau
interaction lenght in Earth Crust, that has been taken into
account. \cite{Fargion2000-2002}, .} \label{fig:fig18}
\end{figure}
\vspace{2.cm}

\section{Summary and Conclusions}
The discover of the expected UHE neutrino Astronomy is urgent and
just behind the corner. Huge volumes are necessary. Beyond
underground $km^3$ detectors a new generation of UHE neutrino
calorimeter lay on front of mountain chains and just underneath
our feet: The Earth itself  offers huge Crown Volumes as Beam Dump
calorimeters observable via upward Tau Air Showers, UPTAUs and
HORTAUs. Their effective Volumes as a function of the quota $h_1$
has been derived by an analytical function variables in equation
$(17), (20),(23),(28)$ and Appendix B. These Volumes are
discussed below and summirized in the last column of next Table
(Fig.9) at the end of the paper. At a few tens meter altitude the
UPTAUs and HORTAUs Ring are almost overlapping. At low altitude
$h_{1}\leq 2 km$ the HORTAUs are nearly independent on the $\rho$
matter density:
${\Delta M_{eff.}(h_{1}=2 km)(\rho_{Water})}= 0.987 km^3 $
${\Delta M_{eff.}(h_{1}=2km)(\rho_{Rock})}= 0.89 km^3
$ These volumes are the effective Masses expressed in Water
equivalent volumes. On the contrary at higher quotas, like highest
Mountain observations sites, Airplanes, Balloons and Satellites,
the matter density of the HORTAUs Ring Areas play a more and more
dominant role asymptotically $proportional$ to ${\rho}^{-2}$:
${\Delta M_{eff.}(h_{1}=5 km)(\rho_{Water})}= 3.64 km^3 $
${\Delta M_{eff.}(h_{1}=5 km)(\rho_{Rock})}= 2.17 km^3 $
From Air-planes or balloons the effective volumes $M_{eff.}$
increases and the density $\rho$ plays a relevant role.
${\Delta M_{eff.}(h_{1}=25 km)(\rho_{Water})}= 20.3 km^3
$
${\Delta M_{eff.}(h_{1}=25 km)(\rho_{Rock})}= 6.3 km^3
$ Finally from satellite altitudes the same effective volumes
$M_{eff.}$ are reaching extreme values.
\begin{displaymath}
{\Delta M_{eff.}(h_{1}=500 km)(\rho_{Water})}= 150.6 km^3
\end{displaymath}
\begin{displaymath}
{\Delta M_{eff.}(h_{1}=500 km)(\rho_{Rock})}=30 km^3
\end{displaymath}
These Masses must be compared with other proposed $km^3$
detectors, keeping in mind that these HORTAUs signals conserve the
original UHE $\nu$ direction information within a degree. One
has  to discriminate  HORTAUS (only while observing from
satellites) from Horizontal High Altitude Showers (HIAS)
\cite{Fargion2001b}, due to rare UHECR showering on high
atmosphere. One might also remind the UPTAUs (at PeVs energies)
volumes as derived in Appendix B and in \cite{Fargion2000-2002}
whose values (assuming an arrival angle$\simeq 45^o- 60^o$ below
the horizons) are nearly $proportional$ to the $\rho$ density:
\begin{displaymath}
{\Delta M_{eff.}(h_{1}=500 km)(\rho_{Water})}= 5.9 km^3
\end{displaymath}
\begin{displaymath}
{\Delta M_{eff.} (h_{1}=500 km)(\rho_{Rock})}=15.6 km^3
\end{displaymath}
These widest Masses values, here estimated analytically for any
quota, are offering an optimal opportunity to reveal UHE $\nu$ at
PeVs and EeVs-GZK energies by crown array detectors
(scintillators, Cherenkov, photo-luminescent) facing vertically
the Horizontal edges, located at high mountain peaks or at
air-plane low sides and finally on  balloons and satellites. As
it can be seen in last column of final Table , the ratio $ R $
between HORTAUs events and Showers over atmospheric UHE $\nu$
interaction is a greater and greater number with growing height,
implying a dominant role (above two order of magnitude) of
HORTAUS grown in Earth Skin Crown over Atmospheric HORTAUs.
\section{Event Rate of Upward and Horizontal Tau Air-showers}
The event rate for HORTAUs are given by the following expression
normalized to any given neutrino flux ${\Phi_{\nu}}$:
\begin{equation}
{\dot{N}_{year}}={\Delta
M_{eff.}}\cdot{\Phi_{\nu}}\cdot{\dot{N_o}}\cdot\frac{\sigma_{E_{\nu
}}}{\sigma_{E_{\nu_o}}}
\end{equation}
Where the ${\dot{N_o}}$ is the UHE neutrino rate estimated for
$km^3$ at any given (unitary) energy ${E_{\nu_o} }$, in absence of
any Earth shadow. In our case we shall normalize our estimate at
${E_{\nu_o}=3}$ PeVs energy for standard electro-weak charged
current in a standard parton model \cite{Gandhi et al 1998} and
we shall assume a  model-independent neutrino maximal flux
${\Phi_{\nu}}$ at a flat fluence value of nearly ${\Phi_{\nu}}_o$
$\simeq 3\cdot 10^3 eV cm^{-2}\cdot s^{-1}\cdot sec^{-1}\cdot
sr^{-1}$ corresponding to a characteristic Fermi power law in UHE
$\nu$ primary production rate decreasing as $\frac
{dN_{\nu}}{dE_{\nu}}\simeq {E_{\nu}}^{-2}$ just below present
AMANDA bounds. The consequent rate becomes:
\begin{displaymath}
{\dot{N}_{year}}= 29 {\frac{{\Delta
M_{eff.}}}{km^{3}}\cdot\frac{{\Phi_{\nu}}}{{\Phi_{\nu}}_o}}\cdot\frac{\sigma_{E_{\nu
}}}{\sigma_{E_{\nu_o}}}
\end{displaymath}
\begin{equation}
=29\cdot{\left(\frac{E_{\nu}}{3 \cdot 10^6 \cdot
GeV}\right)^{-0.637}} {\frac{{\Delta
M_{eff.}}}{km^{3}}\cdot\frac{{\Phi_{\nu}}}{{\Phi_{\nu}}_o}}
\end{equation}
For highest satellites and for a characteristic UHE GZK energy
fluence ${\Phi_{\nu}}_o$  $\simeq 3 10^3 eV cm^{-2}\cdot
s^{-1}\cdot sr^{-1}$(as the needed Z-Showering one), the
consequent event rate observable ${\dot{N}_{year}}$ above the Sea
is :
\begin{equation}
=12.3\cdot{\left(\frac{E_{\nu}}{3 \cdot 10^{10} \cdot
GeV}\right)^{-0.637}} {\frac{{h}}{500
km}\cdot\frac{{\Phi_{\nu}}}{{\Phi_{\nu}}_o}}
\end{equation}
 This event rate is comparable to UPTAUS one and it may be an
 additional source of Terrestrial Gamma Flashes observed by GRO
 in last decade \cite{Fargion2000-2002}.

  In final Fig. 10 it has been  summirized the consequent different detector
  thresholds (at height h= 2,5,25,500 km)
  assuming a ten years records
  over different neutrino flux models
  and in respect with known bounds \cite{Kalashev:2002kx}.

\section{Appendix A}
 As soon as the altitude $h_1$ and the corresponding energy
$E_{\tau_{h_1}}$ increases the corresponding  air density
decreases. At a too high quota there is no more $X$ slant depth
for any Air-Showering to develop. Indeed its value is :
\begin{displaymath}
X=\int_{\frac{d_u}{2}+c\tau\gamma_t}^{d_1+\frac{d_u}{2}}{{n_0}e^{-\frac{R_{\oplus}}{h_0}{\left[{\sqrt{\left(1-\frac{h_2}{R_{\oplus}}\right)^2+\left({\frac{x}{R_{\oplus}}}\right)^2}}-1\right]}}{dx}}
\end{displaymath}
\begin{equation}
\simeq\int_{\frac{d_u}{2}+c\tau\gamma_t}^{d_1+\frac{d_u}{2}}{n_0e^{-\frac{x^2}{2R_{\oplus}h_0}}dx}\leq{n_0h_0}
\end{equation}
\end{enumerate}
In order to find this critical height $h_{1}$ where the maximal
energy HORTAU terminates  we remind our recent approximation. The
transcendental equation that defines the Tau distance $c\tau$
 has been more simplified in:
\begin{equation}\label{13}
  \int_{0}^{+ \infty} n_0 e^{-\frac{\sqrt{(c\tau+x)^2+R_\oplus^2} - R_\oplus}{h_0}}
   dx \cong n_0 h_0 A
\end{equation}
\begin{equation}\label{14}
  \int_{0}^{+ \infty} n_0 e^{-\frac{(c\tau+x)^2} {2h_0R_\oplus}}
   dx \cong n_0 h_0 A
\end{equation}
\begin{equation}\label{15}
 c\tau = \sqrt{2R_\oplus h_0}
 \sqrt{ln \ffrac{R_\oplus}{c\tau} - ln A }
\end{equation}
Here $A=A_{Had.}$ or $A=A_{\gamma}$ are slow logarithmic
functions of values near unity; applying known empirical laws to
estimate this logarithmic growth (as a function of  the X slant
depth) we  derived respectively for hadronic and gamma UHECR
showers \cite{Fargion2000-2002}, \cite{Fargion2001a}:
\begin{equation}\label{5}
 A_{Had.}=0.792 \left[1+0.02523 \ln\ffrac{E}{10^{19}eV}\right]
\end{equation}
\begin{equation}\label{5}
 A_{\gamma}=\left[1+0.04343\ln\ffrac{E}{10^{19}eV}\right]
\end{equation}
The solution of the above transcendental equation leads to a
characteristic maximal UHE $c\tau_{\tau}$ = $546 \;km$ flight
distance, corresponding to $E_1.1\cdot 10^{19}eV$ energy whose
decay occurs at height $H_o= 23$ km; from there on the HORTAUS
begins to shower. At higher quotas the absence of air density
lead to a suppressed development or to a poor particle shower,
hard to be detected. At much lower quota the same air opacity
filter most of the electromagnetic Shower allowing only to muon
bundles to survive at low $(\leq 10^{-3})$ level.
\section{Appendix B: The UPTAUS Area}
The Upward Tau Air-Showers, mostly at PeV energies, might travel a
minimal air depth before reaching the observer in order to
amplify its signal. The UPTAUS Disk Area $A_U$ underneath an
observer at height $h_1$ within a opening angle $\tilde{\theta}_2$
from the Earth Center is:
\begin{equation}
 A_{U}= 2\pi{R_{\oplus}}^2(1 - \cos{\tilde{\theta}_{2}})
\end{equation}
Where the $\sin{\tilde{\theta}_{2}= (x_2/{R_{\oplus}})}$ and
$x_2$ behaves like $x_1$ defined above for HORTAUs. In general the
UPTAUs area are constrained in a narrow Ring (because the mountain
presence itself or because the too near observer distances from
Earth are encountering a too short air slant depth for showering
or a too far and opaque atmosphere for the horizontal UPTAUs):
\begin{equation}
 A_{U}= 2\pi{R_{\oplus}}^2( \cos{\tilde{\theta}_{3}-\cos{\tilde{\theta}_{2}}})
 \end{equation}
An useful Euclidean approximation is:
\begin{equation}
 A_{U}= \pi {h_1}^2 ({\cot{\theta}_{2}}^2-{\cot{\theta}_{3}}^2)
 \end{equation}
 Where ${\theta}_{2}$, ${\theta}_{3}$ are the outgoing $\tau$
 angles on the Earth surface \cite{Fargion 2000-2002}.

 For  UPTAUs (around $3\cdot10^{15} eV$ energies) these
volumes have been estimated in \cite{Fargion 2000-2002}, assuming
an arrival values  angle$\simeq 45^o- 60^o$ below the horizons.
For two characteristic densities one finds respectively:
$${\Delta M_{eff.}(h_{1}=500 km)(\rho_{Water})}= 5.9~ km^3;$$
$${\Delta M_{eff.}.  (h_{1}=500 km)(\rho_{Rock})}=15.6~ km^3 $$
Their detection efficiency is displayed in last Figure , and it
exceed by more than an order of magnitude, the future ICE-CUBE
threshold.

\subsection*{Acknowledgements} \leftskip = 0cm \rightskip = 0cm
 The author wishes to thank  P.G.De Sanctis Lucentini, C.Leto, M.De Santis
 for  numerical and technical support and Prof.G.Salvini and Prof. B.Mele for useful
 discussions and comments.

\end{document}